\documentclass{JHEP3}
\usepackage{epsfig,amsmath,bbm}
\usepackage{dsfont}
\usepackage{wasysym}
\usepackage{cite}

\newcommand{\be}{\begin{equation}}
\newcommand{\ee}{\end{equation}}
\newcommand{\bea}{\begin{eqnarray}}
\newcommand{\eea}{\end{eqnarray}}
\newcommand{\nn}{\nonumber}
\newcommand{\ba}{\begin{array}} 
\newcommand{\ea}{\end{array}}
\newcommand{\lmat}[2][lllll]{\left( \begin{array}{#1} #2\\ \end{array} \right)}

\newcommand{\rmat}[2][rrrrr]{\left( \begin{array}{#1} #2\\ \end{array} \right)}
\newcommand{\lsim}{
\mathrel{\hbox{\rlap{\hbox{\lower4pt\hbox{$\sim$}}}\hbox{$<$}}}}
\newcommand{\gsim}{
\mathrel{\hbox{\rlap{\hbox{\lower4pt\hbox{$\sim$}}}\hbox{$>$}}}}

\newcommand{\svev}{\left<S\right>}

\newcommand{\md}{\text{~mod~}}
\newcommand{\util}{\widetilde u}
\newcommand{\dtil}{\widetilde d}
\newcommand{\etil}{\widetilde e}
\newcommand{\ntil}{\widetilde n}

\newcommand{\bs}{\boldsymbol}

\title{
\vspace*{.25in}
Common gauge origin of discrete symmetries
\\in observable sector and hidden sector
}

\author{Taeil Hur\\
Department of Physics, KAIST, Daejon 305-701, Korea\\
E-mail: \email{tailor@muon.kaist.ac.kr}}
\author{Hye-Sung Lee\\
Institute for Fundamental Theory, University of Florida, Gainesville, FL 32611, USA\\
Department of Physics and Astronomy, University of California, Riverside, CA
92521, USA \\
E-mail: \email{hlee@phys.ufl.edu}}
\author{Christoph Luhn\\
Institute for Fundamental Theory, University of Florida, Gainesville, FL 32611, USA\\
School of Physics and Astronomy, University of Southampton, Highfield, Southampton, SO17 1BJ, UK\\
E-mail: \email{christoph.luhn@soton.ac.uk}}

\preprint{UFIFT-HEP-08-17\\UCRHEP-T457\\SHEP-08-32}

\abstract{
An extra Abelian gauge symmetry is motivated in many new physics models in both supersymmetric and nonsupersymmetric cases.
Such a new gauge symmetry may interact with both the observable sector and the hidden sector.
We systematically investigate the most general residual discrete symmetries in both sectors from a common Abelian gauge symmetry.
Those discrete symmetries can ensure the stability of the proton and the dark matter candidate.
A hidden sector dark matter candidate (lightest $U$-parity particle or LUP) interacts with the standard model fields through the gauge boson $Z'$ which may selectively couple to quarks or leptons only.
We make a comment on the implications of the discrete symmetry and the leptonically coupling dark matter candidate, which has been highlighted recently due to the possibility of the simultaneous explanation of the DAMA and the PAMELA results.
We also show how to construct the most general $U(1)$ charges for a given discrete symmetry, and discuss the relation between the $U(1)$ gauge symmetry and $R$-parity.
}


\begin{document}

\section{Introduction}
\label{sec:introduction}
For many beyond standard models, discrete symmetries are invaluable ingredients to make the models phenomenologically viable.
For example, in the minimal supersymmetric standard model (MSSM), $R$-parity
\cite{Farrar:1978xj} is usually assumed for the proton stability.
$R$-parity also guarantees the stability of the lightest superparticle (LSP), which can be a good dark matter candidate.
It is argued, however, that discrete symmetries are vulnerable to  Planck scale physics unless they have a gauge origin \cite{Krauss:1988zc}.
An extra Abelian gauge symmetry is also predicted in many new physics scenarios such as superstring, extra dimension, little Higgs, and grand unification.
Therefore, it would be useful to understand what discrete symmetries are allowed as a residual discrete symmetry of the extra $U(1)$ gauge symmetry.

The first systematic study of the $U(1)$ residual discrete symmetry in a supersymmetry (SUSY) framework was performed by Ibanez and Ross \cite{Ibanez:1991pr}, where they found 3 independent generators $R_N$, $L_N$, and $A_N$.
They studied all possible $Z_2$ and $Z_3$ discrete symmetries from a $U(1)$, and found that $R_2$ (matter parity, which is equivalent to $R$-parity) as well as another $Z_3$ symmetry can be a residual discrete symmetry of the gauge symmetry, a.k.a. a discrete gauge symmetry.
Complementary and general discrete symmetries ($Z_N$ with $N > 3$) with a $U(1)$ origin were also studied \cite{Dreiner:2005rd,Luhn:2007gq}.
In a special case where the $\mu$-problem \cite{Kim:1983dt} is addressed by a TeV scale $U(1)$, the discrete symmetries were investigated in Refs.~\cite{Lee:2007fw,Lee:2007qx}, which allow $R$-parity violating $U(1)$ models without fast proton decay.

Nevertheless, these discrete symmetries concerned only the observable sector (or the MSSM sector).
Many theories need exotic chiral fields for various reasons.
For example, the SUSY breaking mechanism requires additional  fields.
Also  exotic fields are often necessary to make the model anomaly free when an additional gauge symmetry is added.
Even when they do not have standard model (SM) charges, such hidden sector fields may have charges under the extra $U(1)$ gauge symmetry.
The SM neutral hidden sector fields can be natural dark matter candidates if they are stable.

It was shown that the same $U(1)$ symmetry that provides the discrete symmetry for the MSSM sector can also be the source of the discrete symmetry for the hidden sector simultaneously \cite{Lee:2008pc}.
Another independent generator $U_N$ was introduced for the hidden sector discrete symmetry.
The lightest $U$-parity particle (LUP) from the hidden sector is stable under the $U_2$ ($U$-parity), and it was shown that the experimental constraints from the relic density and the direct detection can be satisfied in a large parameter space with the LUP dark matter candidate \cite{Hur:2007ur}.

However, the study in Ref.~\cite{Lee:2008pc} was not completely general since the hidden sector field was assumed to be Majorana with $SXX$ as a mass term, and only the factorizable extension $Z_{N_1}^{\rm obs} \times Z_{N_2}^{\rm hid}$ was exploited.
In this paper, we first generalize the discussion by including the Dirac type hidden sector fields and possible nonrenormalizable mass terms.
Dirac type fields allow a discrete symmetry $U_N$ (with $N>2$) while Majorana type fields allow only~$U_2$.
This leads to the possibility of multiple hidden sector dark matter candidates stable due to the hidden sector discrete symmetry.
We also start from the general form of the discrete symmetry taking the factorizable case as a special limit.
Then we present a method to construct the most general $U(1)$ charges for a given discrete symmetry of the MSSM and hidden sector, with illustrations for specific examples.
In Appendix \ref{sec:RN}, we discuss the $U(1)$ origin of the popular $R$-parity and its relation with the $U(1)$ solution of the $\mu$-problem, which is one of the motivations to extend the supersymmetric standard model to include an extra $U(1)$.
In Appendix \ref{sec:leptonic}, we discuss about the compatibility of discrete symmetries with a leptonically coupling dark matter candidate.

\section{Residual discrete symmetries from the $\bs{U(1)}$ gauge symmetry}
\label{sec:discrete}
In this section, we review the general discrete symmetries in the MSSM sector, which are the remnant of an Abelian gauge symmetry.
Starting with a $U(1)$ gauge symmetry which is broken spontaneously by a Higgs singlet $S$, one is generically left with a residual discrete $Z_N$ symmetry.
In a normalization where all particles $F_i$ of the theory have integer $U(1)$ charges $z[F_i]$, the value of $N$ is directly determined by 
\be
N = |z[S]| \ . \label{eq:zS}
\ee
The resulting discrete charges $q[F_i]$ of the fields $F_i$ are then given by the $\md N$ part of their original $U(1)$ charges
\be
q[F_i] = z[F_i] \md N \ . \label{eq:qF}
\ee
By definition the Higgs singlet $S$ has vanishing discrete charge so that giving a vacuum expectation value (vev) to $S$ keeps the discrete $Z_N$ symmetry unbroken.
Note that in the case with $N=1$, we formally obtain a $Z_1$ which corresponds to no remnant discrete symmetry.

$\!$The possible (family-independent) discrete symmetries of the MSSM\footnote{We include 3 right-handed neutrinos $N^c$ which do not change our argument.} which can emerge from an anomaly free $U(1)$ gauge symmetry have been identified and investigated in Refs.~\cite{Ibanez:1991pr,Dreiner:2005rd,Luhn:2007gq}.
Demanding $Z_N$ invariance of the MSSM superpotential operators
\bea
W_{\mu} &=& \mu H_u H_d \ , \label{eq:muterm}\\
W_{\rm Yukawa} &=& y^D_{jk} H_d Q_j D^c_k + y^U_{jk} H_u Q_j U^c_k + y^E_{jk}
H_d L_j E^c_k + y^N_{jk} H_u L_j N_k^c \ , \label{eq:yukawa}
\eea
one can express any discrete symmetry {\it among the MSSM particles} in terms of the two generators 
\be
R_N = e^{2 \pi i (q_R/N)} \ , \qquad L_N = e^{2 \pi i (q_L/N)} \ ,
\ee
where the charges $q_R$ and $q_L$ are defined in Table~\ref{tab:discrete}.
Different discrete symmetries of the observable sector are then obtained by multiplying various integer powers of these generators
\be
g^{\rm obs}_N = R_N^m L_N^p \ . \label{eq:gvis}
\ee 
Compared to Refs.~\cite{Ibanez:1991pr,Dreiner:2005rd}, the generator $A_N$, which gives nonzero discrete charge to only one of the two Higgs doublets, is omitted because its presence would forbid the $\mu$ term in Eq.~\eqref{eq:muterm}.
As the invariance of $H_uH_d$ under $Z_N$ requires opposite discrete charges for $H_u$ and $H_d$, one can always find an equivalent set of discrete charges by adding some amount of hypercharge $y[H_i]$ such that $q'[H_i] = q[H_i] + \alpha y[H_i] = 0$ simultaneously for $i=u$, $d$.
Thus requiring the existence of the $\mu$ term guarantees the absence of domain walls after the electroweak symmetry breaking.

A more intuitive way of writing Eq.~\eqref{eq:gvis} is obtained by defining the generator $B_N = R_N L_N$.
The discrete charges $q_B$ of the MSSM fields under $B_N$ are related to the familiar baryon number ($\cal B$) by the hypercharge shift 
\be
q_B[F_i] =  -{\cal B}[F_i] + \frac{1}{3} y[F_i] \ .
\ee
Here the hypercharge is normalized so that $y[Q]=1$. On the other hand, the discrete charges~$q_L$ of the MSSM fields under $L_N$ are nothing but the
negative of the lepton number ($\cal L$)
\be
q_L[F_i] = -{\cal L}[F_i] \ .
\ee
Hence, the general discrete symmetry of Eq.~\eqref{eq:gvis}, written in terms of $B_N$ and $L_N$, 
\be
Z_N^{\rm obs} ~:~ g^{\rm obs}_N ~=~ B_N^b L_N^\ell \ , \label{eq:gvisBL}
\ee
can be understood in terms of the well-known baryon number and lepton number, with a discrete charge
\be
q = b q_B + \ell q_L \md N = - b {\cal B} -\ell {\cal L} + b (y/3) \md N \ .
\ee
The exponents in Eqs.~\eqref{eq:gvis} and \eqref{eq:gvisBL} are related to each other by $m=b$ and $p=b+\ell$. Specific values for $b$ and $\ell$ define a $Z_N$ symmetry of the MSSM particles for which the quantity ${\cal Q} = b {\cal B} + \ell {\cal L} \md N$ is conserved.
The lightest particle with nonzero $\cal Q$ value will be stable by the discrete symmetry.
The following discrete symmetries are some examples obtained for given $b$ and $\ell$ values.
\be
\begin{array}{ccc}
\quad(b,\ell;N)\quad & \qquad g_N^{\rm{obs}} \qquad & \qquad {\cal Q} \qquad
\\[1mm]\hline\hline 
(1,0;N) & B_N & {\cal B} \md N  \\[1mm]
(0,1;N) & L_N & {\cal L} \md N  \\[1mm]
(1,1;N) & B_N L_N & ({\cal B} + {\cal L}) \md N \\[1mm]
(1,-1;N) & B_N L_N^{-1} & ({\cal B} - {\cal L}) \md N
\end{array} \nn
\ee
Note that with $N=2$ the symmetry in the last line ($B_2 L_2^{-1}$) corresponds to matter parity because $(-1)^{{\cal B} - {\cal L}} = (-1)^{3({\cal B} - {\cal L})}$ for any $SU(3)_C$ invariant term for which $\cal B$ is always an integer.
As long as the spin angular momentum is conserved, matter parity is equivalent to $R$-parity, $R_p=(-1)^{3({\cal B} - {\cal L}) + 2s}$. 

The discussion so far has been completely independent of any assumptions about the origin of the discrete symmetry.
Requiring that the $Z_N$ arises as a remnant of an anomaly free $U(1)$ gauge symmetry, we have to impose the discrete anomaly conditions of Ref.~\cite{Ibanez:1991pr} (Note the cubic anomaly condition is disregarded \cite{Banks:1991xj}.)
\bea
{[SU(3)_C]}^2 - U(1) &:& \sum_{i={\bf 3},\overline {\bf 3}} q_i = N \cdot {\bf Z} \ , \label{eq:dac1} \\
{[SU(2)_L]}^2 - U(1) &:& \;\sum_{i={\bf 2}} ~q_i = N \cdot {\bf Z} \ , \label{eq:dac2} \\
{[\text{gravity}]}^2 - U(1) &:&  \;\sum_i ~q_i =\left\{ 
\begin{array}{ll}N \cdot {\bf Z} & ~(N=\text{odd}) \ , \\[1mm]
  \frac{N}{2} \cdot {\bf Z} & ~(N=\text{even}) \ ,
\end{array}\right. \label{eq:dac3} 
\eea
where the sums run over MSSM particles only.
Additional exotic fields which may or may not be singlets under the SM gauge group $SU(3)_C \times SU(2)_L \times U(1)_Y$ do not contribute to these anomaly conditions as long as they acquire a mass term when $U(1)$ is broken, i.e. if they are vectorlike under the SM gauge groups, while they are not under the $U(1)$.

The consequence of Eqs.~\eqref{eq:dac1} - \eqref{eq:dac3} is that some sets of parameters $(b,\ell ;N)$ correspond to $Z_N$ symmetries which are
discrete anomaly free while others are anomalous and therefore ruled out (see Refs.~\cite{Ibanez:1991pr,Dreiner:2005rd,Luhn:2007gq}).
For instance, the symmetries of type $B_N^b$ automatically satisfy Eqs.~\eqref{eq:dac1} and \eqref{eq:dac3} for all $N$ and $b$.
However, Eq.~\eqref{eq:dac2} yields the nontrivial constraint
\bea
\sum_{i={\bf 2}} q_i &=& b \left\{ N_f (3 q_B[Q]+q_B[L])+N_H (q_B[H_u]+q_B[H_d]) \right\}  \\
&=& - b N_f ~=~ 0 \md N \ , \label{eq:NfN}
\eea
where $N_f$ and $N_H$ denote the number of families of fermions and Higgs pairs, respectively.
For $N_f=3$, we obtain only $b=0$, $\pm N/3$ as allowed choices for the cyclic symmetry.
Unless $b=0$ (which is not a real symmetry), we are led to the symmetry $B_{|3 b|}^{b}$ or $B_3$.
Similarly, the discrete anomaly free symmetries of type $L_N^\ell$ are only $L_{|3\ell|}^\ell$ or $L_3$ (unless $\ell = 0$).
This conclusion does not depend on whether there are massive $SU(2)_L$ charged exotics or not since their corresponding mass term would imply vanishing contribution to the discrete anomaly condition \cite{Ibanez:1991hv}.

\TABLE{
\caption{
Discrete charges of $R_N$, $B_N$, $L_N$, $U_{a,N}$, $U'_{a,N}$ and their relation with $\cal B$, $\cal L$, ${\cal U}_a$, and ${\cal U}'_a$.
\label{tab:discrete}}
{\small
\begin{tabular}{|l|c||r|r|r|r|r|r|r|r|r|r|r|l|c|}
\hline
\multicolumn{2}{|c||}{symmetry}    & $\!Q\!$   & $\!\!U^c\!\!$ & $\!\!D^c\!\!$ &
$\!L\!$   & $\!\!E^c\!\!$ &$\!\!N^c\!\!$ & $\!\!H_u\!\!$ & $\!\!H_d\!\!$ & $\!X_b\!$ &
$\!T_b\!$ & $\!\!T_b^c\!\!$   & $\!$meaning of $q\!$   & conserved $\cal{Q}$ \\
\hline\hline
$\!R_N$ & $q_R$         & $\!0\!$   & $\!\!\!-1\!$  & $\!1\!$   & $\!0\!$   &
$\!1\!$   & $\!\!\!-1\!$  & $\!1\!$ &$\!\!\!-1\!$  & $\!0\!$ & $\!0\!$ & $\!0\!$
& $\! y/3 -({\cal B}\!-\!{\cal L})\!\! $ &  $\!\!({\cal B}\!-\!{\cal L}) \!\md \! N\!\!$ \\ 
& $3q_R - y$& $\!\!\!-1\!$  & $\!1\!$   & $\!1\!$   & $\!3\!$   & $\!\!\!-3\!$
& $\!\!\!-3\!$  & $\!0\!$   & $\!0\!$   & $\!0\!$ & $\!0\!$ & $\!0\!$     &
$\!-3({\cal B}\!-\!{\cal L})\!$ & \\
\hline
$\!B_N$ & $q_B$         & $\!0\!$   & $\!\!\!-1\!$  & $\!1\!$   & $\!\!\!-1\!$
&$\!2\!$   & $\!0\!$   & $\!1\!$   & $\!\!\!-1\!$  & $\!0\!$ & $\!0\!$ &
$\!0\!$ & $\!y/3-{\cal B} $ &      ${\cal B} \md N$ \\ 
&$3q_B-y$& $\!\!\!-1\!$  & $\!1\!$   & $\!1\!$   & $\!0\!$   & $\!0\!$   &
$\!0\!$   & $\!0\!$   & $\!0\!$ & $\!0\!$ & $\!0\!$ & $\!0\!$     & $\!-3{\cal
  B}$     & \\ 
\hline
$\!L_N$ &  $q_L$      & $\!0\!$   & $\!0\!$   & $\!0\!$   & $\!\!\!-1\!$  &
$\!1\!$   & $\!1\!$   & $\!0\!$   & $\!0\!$   & $\!0\!$ & $\!0\!$ & $\!0\!$
& $\!-{\cal L}$      & ${\cal L} \md N$ \\ 
\hline
$\!U_{a,N}\!\!$ & $q_{U_a}$ & $\!0\!$ & $\!0\!$ & $\!0\!$ & $\!0\!$ & $\!0\!$
& $\!0\!$ & $\!0\!$ & $\!0\!$ & $\!\!\!-\delta_{ab}\!\!$ & $\!0\!$ & $\!0\!$ &
$\!-{\cal U}_a$      & ${\cal U}_a \md N$ \\ 
\hline
$\!U'_{a,N}\!\!$ & $q_{{U'}_{\!\! a}}$ & $\!0\!$ & $\!0\!$ & $\!0\!$ & $\!0\!$
& $\!0\!$ & $\!0\!$ & $\!0\!$ & $\!0\!$ & $\!0\!$ & $\!\!\!-\delta_{ab}\!\!$ &
$\!\!\delta_{ab}\!\!$ & $\!-{\cal U}'_a$ & ${\cal U}'_a \md N$ \\
\hline
\hline
\multicolumn{2}{|l||}{$\!y\!$ (hypercharge)$\!\!$}           & $\!1\!$   &
$\!\!\!-4\!$ & $\!2\!$   & $\!\!\!-3\!$  & $\!6\!$   & $\!0\!$   & $\!3\!$   &
$\!\!\!-3\!$  & $\!0\!$    & $\!0\!$  &  $\!0\!$     & &  \\
\hline
\multicolumn{2}{|l||}{$\!{\cal B}\!$ (baryon no.)$\!$}    & $\!\frac{1}{3}\!$
& $\!\!\!-\frac{1}{3}\!$  & $\!\!\!-\frac{1}{3}\!$   & $\!0\!$  & $\!0\!$   &
$\!0\!$   & $\!0\!$   & $\!0\!$  & $\!0\!$    &    $\!0\!$  &  $\!0\!$     &
&  \\ 
\hline
\multicolumn{2}{|l||}{$\!{\cal L}\!$ (lepton no.)$\!$}    & $\!0\!$   & $\!0\!$
& $\!0\!$   & $\!1\!$  & $\!\!\!-1\!$   & $\!\!\!-1\!$   & $\!0\!$   & $\!0\!$
& $\!0\!$    &   $\!0\!$   &    $\!0\!$   &                  & \\ 
\hline
\end{tabular}
}
}

\section{Hidden sector discrete symmetries}
We now wish to extend the concept of discrete symmetries to the hidden sector or the SM neutral particles.\footnote{The discrete symmetry argument does not change even if the Dirac type exotics are SM-charged.}
To do so, we introduce the generators $U_{a,N}$ and $U'_{a,N}$ which assign nontrivial discrete charges to, respectively, the Majorana ($X_b$) and the Dirac ($T_b$, $T_b^c$) particles of the hidden sector while the MSSM fields remain uncharged.
Note that we label hidden sector fields by indices ($a$, $b$, etc) which can refer to fields with different or identical (i.e. family) $U(1)$ charges.

The generators $U_{a,N}$, $U'_{a,N}$ as well as $R_N$, $L_N$, and $B_N$ -- extended to include the hidden sector particles -- are shown in Table~\ref{tab:discrete}.
Introducing the discrete symmetry of the hidden sector
\be
g_N^{\rm{hid}} = U_N = \prod_{a,b} U_{a,N}^{u_a} \, U_{b,N}^{\prime u'_b} \ ,
\ee
the generalized discrete symmetry over the observable and the hidden sectors can be written as  
\be
Z_N ~:~ g_N = g_N^{\rm{obs}} g_N^{\rm{hid}} = B_N^b L_N^\ell U_{N} \ . \label{eq:ext-dsym}
\ee
It is uniquely determined by the integer exponents $(b,\ell,u_a,u'_b;N)$, entailing the discrete charges 
\be
q = b q_B + \ell q_L + u_a q_{U_a} + u'_b q_{U'_b}  ~\md N \ .
\ee
Summation over repeated indices is assumed as usual.
Under the assumption that the hidden sector particles acquire a mass after the gauge symmetry $U(1)$ is broken down to the discrete symmetry, invariance of the bilinear terms
\be
W_{\rm{hidden}} = m_{a} X_a X_a + m'_{b} T_b T^c_b \label{eq:hidmass} 
\ee
under $Z_N$ constrains the exponents $u_a$ to
\be
u_a =0 ~~\text{or}~~ N/2 \ ,
\ee
which makes it effectively a $Z_2$ parity for Majorana type field ($X$).

Starting with an anomaly free discrete symmetry $g_N^{\rm{obs}}$ in the observable sector, the extended  discrete symmetry $g_N$ can also originate in an anomaly free $U(1)$ gauge symmetry, regardless of the chosen values for $u_a$ and $u'_b$.
In other words, due to the $Z_N$ invariance of the mass terms in Eq.~\eqref{eq:hidmass}, $g_N$ and $g_N^{\rm{obs}}$ {\it jointly} either satisfy or do not satisfy the discrete anomaly conditions of Eqs.~\eqref{eq:dac1} - \eqref{eq:dac3}.
Now we consider the case where $Z_N$ can be factorized into two smaller discrete symmetries.
\be
U(1) ~ \rightarrow ~ Z_N = Z_{N_1} \times Z_{N_2} \ ,
\ee
where $N=N_1 N_2$.
This decomposition is only possible if $N_1$ and $N_2$ have no common prime factor, i.e. they must be coprime to each other.
Let us apply this method to separate the discrete symmetries of the observable and the hidden sector.
To do so, we have to assume that the exponents $b$ and $\ell$ are multiples of $N_2$, while $u_a$ and $u'_b$ are multiples of $N_1$.
Eq.~\eqref{eq:ext-dsym} can then be written as
\be
g_N =  B_{N_1N_2}^{b} L_{N_1N_2}^{\ell} \, \prod_{a,b} \, U_{a, N_1N_2}^{u_a} U_{b, N_1N_2}^{\prime u'_b}
= B_{N_1}^{b/N_2} L_{N_1}^{\ell/N_2} \, \prod_{a,b} \, U_{a, N_2}^{u_a/N_1} U_{b, N_2}^{\prime u'_b/N_1} \ .
\ee
This yields a $Z^{\rm{obs}}_{N_1}$ symmetry in the observable sector and a $Z^{\rm{hid}}_{N_2}$ in the hidden sector with charges
\bea
q^{\rm{obs}}_{Z_{N_1}} &=& \left(\frac{b}{N_2}\right) q_B + \left(\frac{\ell}{N_2}\right) q_L ~\md N_1 \ , \\
q^{\rm{hid}}_{Z_{N_2}} &=& \left(\frac{u_a}{N_1}\right) q_{U_a} + \left(\frac{u'_b}{N_1}\right) q_{U'_b} ~\md N_2 \ . \label{eq:qZN2hid}
\eea
Both originate in the underlying $Z_N$ symmetry and are conserved separately.
The symmetry $Z^{\rm{obs}}_{N_1}$ can be used to forbid certain processes whose external states comprise only MSSM particles.
On the other hand, the $Z^{\rm{hid}}_{N_2}$ symmetry can stabilize the lightest $\cal U$ charged particle, leading to a dark matter candidate in the hidden sector \cite{Hur:2007ur, Lee:2008pc}.

Depending on $N_2$ as well as the $Z_{N_2}$ charges $q^{\rm{hid}}_{Z_{N_2}}$, there could be even more than one hidden sector particle stable due to the discrete symmetry.
Assume  that $N_2 = \prod_k \, n_k$, where all factors $n_k$ are coprime to each other.
Evidently, all but perhaps one $n_k$ are necessarily odd. Then, the decomposition of the discrete symmetry in the hidden sector reads  
\be
Z^{\rm{hid}}_{N_2} = Z^{\rm{hid}}_{n_1} \times Z^{\rm{hid}}_{n_2} \times \cdots \ . 
\ee
What are the charges of the particles $X_a$ and $T_b$ under these individual $Z^{\rm{hid}}_{n_k}$?
Due to the invariance of the mass term for a Majorana particle $X_a$, its $Z^{\rm{hid}}_{n_k}$ charge must be zero for odd $n_k$.
In the case where there is an even $n_k$, the particle $X_a$ has charge 
\be
q^{}_{Z_{n_k}} [X_a] = -\frac{u_a}{N_1} \cdot \frac{n_k}{N_2} = 0 ~~\text{or}~~ \frac{n_k}{2} \quad (n_k = \text{even}) \ ,
\ee
under $Z^{\rm{hid}}_{n_k}$. 
For Dirac particles $T_b$, the $Z^{\rm{hid}}_{n_k}$ charges
$q^{}_{Z_{n_k}} [T_b]$ are related by 
\be
- \frac{u'_b}{N_1} = \sum_k  q^{}_{Z_{n_k}} [T_b]  \cdot \frac{N_2}{n_k}  \md N_2 \ .
\label{eq:DiracSub}
\ee
Since all $n_k$ are coprime to each other, the charges $q^{}_{Z_{n_k}} [T_b]$ are uniquely fixed by the value of $-\,{u'_b}/{N_1}$.\footnote{If there were a
  second charge assignment $\tilde q^{}_{Z_{n_k}} [T_b]$ for the same value of $- {u'_b}/{N_1}$, the sum $\sum_k (q^{}_{Z_{n_k}} [T_b] - \tilde q^{}_{Z_{n_k}} [T_b])/n_k$ would have to be integer. This however is only possible for $q^{}_{Z_{n_k}} [T_b] - \tilde q^{}_{Z_{n_k}} [T_b] =0$.} 
Consider for example three particles $X$, $T_1$, $T_2$, which have the $Z_{60}$ charges $q^{}_{Z_{60}} [X]=30$, $ q^{}_{Z_{60}} [T_1]=24$, $ q^{}_{Z_{60}} [T_2]=35$, respectively.
The $Z_{60}$ symmetry breaks up into  $Z_{4} \times Z_{3}\times Z_{5}$, leading to the following charges. 
\be
\begin{array}{ccccc}
&  q[X] & q[T_1] & q[T_2] & 60 / n_k\\ \hline\hline
Z_4 & 2 & 0 & 1 &15 \\[1mm]
Z_3 & 0 & 0 & 1 &20 \\[1mm]
Z_5 & 0 & 2 & 0 &12 \\[1mm] \hline
Z_{60}  & 30 & 24 & 35 & -
\end{array} \nn
\ee

From Eqs.~\eqref{eq:qZN2hid} and \eqref{eq:DiracSub}, the $Z_{60}^{\rm hid}$ discrete charge for $T_2$, for example, can be written as
\be
q_{Z_{60}}^{\rm hid} [T_2] = 35 = - \frac{u'_b}{N_1} \md 60 = 1 \cdot 15 + 1 \cdot 20 + 0 \cdot 12 \md 60 \ .
\ee

$T_2$ is the only particle charged under the $Z_3$ symmetry.
Thus it is stable.
Similarly $T_1$ is stable because it is the only $Z_5$ charged particle.
Finally, the symmetry $Z_4$ stabilizes the lighter of the two particles $X$ and $T_2$.
If this is $T_2$, then there is no more particle stable due to the discrete symmetry.
In that way, it is possible that different $Z_{n_k}$ symmetries stabilize the same particle.

The important point in this discussion is that {\it a single} $U(1)$ gauge symmetry can effectively give rise to more than one discrete symmetry.
One part of it might be used to forbid unwanted processes involving the MSSM fields only, while other parts lead to stable hidden sector particles, i.e. multiple dark matter candidates.\footnote{Of course, we can have multiple dark matter candidates from the MSSM sector and hidden sector for $Z_N = R_2 \times U_3$, for example, which can provide the LSP dark matter (stable under $R$-parity) and the Dirac type hidden sector dark matter (stable under $U_3$).}
This setup is schematically sketched in Figure~\ref{fig:diagram}.
The discussion here is basically a generalization of that of Ref.~\cite{Lee:2008pc}, which dealt with only the Majorana case with a specific $SXX$ mass term.

An example of the purely hidden sector discrete symmetry in the non-SUSY case can be found in Ref.~\cite{Kubo:2006rm}, where an additional $U(1)$ was introduced to explain the neutrino mass and dark matter simultaneously.

\FIGURE{
\begin{tabular}{|c|c|c|}
\multicolumn{3}{c}{$U(1) ~ \rightarrow ~ Z_N^{\rm{tot}} = Z_{N_1}^{\rm{obs}} \times
Z_{N_2}^{\rm{hid}}$} \\[4mm]
\cline{1-1}\cline{3-3} ~\hspace{30mm}~ &  ~\hspace{10mm}~ &
~\hspace{30mm}~\\[-2.5mm]
MSSM sector & & Hidden sector \\[3mm] 
\cline{2-2} && \\[-5mm]
$Z_{N_1}^{\rm{obs}}~:~ B_{N_1}^b L_{N_1}^{\ell}$ && $Z_{N_2}^{\rm{hid}}~:~ U_{N_2}$ \\[3mm]
\cline{1-1}\cline{3-3}
\end{tabular}
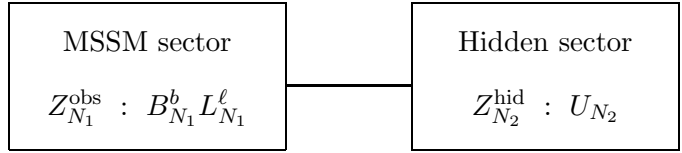
\caption{A unified picture of a single U(1) gauge symmetry that provides the discrete symmetries for the observable sector and the hidden sector.
}
\label{fig:diagram}
}

\section{General ${\bs{U(1)}}$ charges}
\label{sec:formalism}
Having discussed the most general $Z_N$ symmetries that can arise from a $U(1)$ gauge symmetry, we now want to derive the most general $U(1)$ charges within our setup.
Including the possibility that the superpotential terms of Eqs.~\eqref{eq:muterm}, \eqref{eq:yukawa}, and \eqref{eq:hidmass} originate from higher-dimensional operators, the underlying theory before $U(1)$ breaking generally includes the following terms\footnote{In addition to the factors $\left( \frac{S}{M} \right)$ one could also have powers of $\left( \frac{H_u H_d}{M^2} \right)$ multiplying the effective superpotential terms. For the sake of clarity, we omit this possibility.} 
\bea
\hat W_{\mu} &=& \hat \mu \left(\frac{S}{M}\right)^p S H_u H_d \ , \\
\hat W_{\rm Yukawa} &=& 
\hat y^D_{jk} \left(\frac{S}{M}\right)^{\tilde d} H_d Q_j D^c_k 
~+~ \hat y^U_{jk} \left(\frac{S}{M}\right)^{\tilde u}  H_u Q_j U^c_k ~+~\nonumber \\ &&
\!\!\!\!\!+~ \hat y^E_{jk} \left(\frac{S}{M}\right)^{\tilde e}  H_d L_j E^c_k 
~+~ \hat y^N_{jk} \left(\frac{S}{M}\right)^{\tilde n}  H_u L_j N_k^c  \ ,\\
\hat W_{\rm{hidden}} &=& 
\hat m_{a}  \left(\frac{S}{M}\right)^{\tilde x_a} S X_a X_a 
~+~ \hat m'_{b} \left(\frac{S}{M}\right)^{\tilde t_b} S T_b T^c_b \ ,
\eea
where we assume generation independent integer exponents with $0\leq \tilde d, \tilde u, \tilde e, \tilde n$ and $-1 \leq p, \tilde x_a, \tilde t_b$. $M$ is
some high mass scale (e.g. $M_{\rm{GUT}}$ or $M_{\rm{Pl}}$) at which new physics generates the nonrenormalizable operators. Note that $\hat
\mu$, $\hat m_a$, and $\hat m'_b$ are dimensionless parameters.  

These terms yield severe constraints on the allowed $U(1)$ charges of the chiral matter fields. We find
\bea
Y_S&:& (1+p) z[S] + z[H_u] + z[H_d]    = 0 \ ,\label{eq:YS} \\
Y_D&:& z[H_d] + z[Q] + z[D^c] + \dtil z[S] = 0 \ , \label{eq:YD} \\
Y_U&:& z[H_u] + z[Q] + z[U^c] + \util z[S] = 0 \ ,\label{eq:YU} \\
Y_E&:& z[H_d] + z[L] + z[E^c] + \etil z[S] = 0 \ ,\label{eq:YE} \\
Y_N&:& z[H_u] + z[L] + z[N^c] + \ntil z[S] = 0 \ ,\label{eq:YN} \\
Y_{X_a}&:& (1+\tilde x_a) z[S] + 2 z[X_a] = 0 \ ,\label{eq:YX} \\
Y_{T_b}&:& (1+\tilde t_b) z[S] + z[T_b] + z[T_b^c] =0 \ .
\eea 
From this we obtain the general solution of $U(1)$ charges in terms the continuous real parameters $\alpha$, $\beta$, $\gamma$, $\delta$, $\tau_b$
\be
\lmat{
\!\!z[Q]\!\!\\
\!\!z[U^c]\!\!\\
\!\!z[D^c]\!\!\\
\!\!z[L]\!\!\\
\!\!z[E^c]\!\!\\
\!\!z[N^c]\!\!\\
\!\!z[H_u]\!\!\\
\!\!z[H_d]\!\!\\
\!\!z[S]\!\!\\
\!\!z[X_a]\!\!\\
\!\!z[T_b]\!\!\\
\!\!z[T_b^c]\!\!}
\!\!=
\frac{\alpha}{3} \rmat{
\!\! 1\!\\
\!\!\!\!-1\!\\
\!\!\!\!-1\!\\
\!\!\!\!-3\!\\
\!\! 3\!\\
\!\! 3\!\\
\!\! 0\!\\
\!\! 0\!\\
\!\! 0\!\\
\!\! 0\!\\
\!\! 0\!\\
\!\! 0\!}
\!+
\frac{\beta}{6} \rmat{
 \!\!1\!\\
\!\!\!\!-4\!\\
\!\! 2\!\\
\!\!\!\!-3\!\\
\!\! 6\!\\
\!\! 0\!\\
\!\! 3\!\\
\!\!\!\!-3\!\\
\!\! 0\!\\
\!\! 0\!\\
\!\! 0\!\\
\!\! 0\!}
+\frac{\gamma}{3 N_f} \rmat{
 N_H (1+p)\!\\
\!\!\!\!-3N_f\util + (3N_f-N_H) (1+p)\!\\
-3N_f\dtil - N_H (1+p)\!\\
 0\!\\
-3N_f\etil\!\\
-3N_f\ntil+3N_f(1+p)\!\\
-3N_f (1+p)\!\\
 0\!\\
 3N_f\!\\
-3N_f (1+\tilde x_a) /2\!\\
 0\!\\
 -3N_f(1+\tilde t_b)\!}
-\frac{\delta}{3N_f} \rmat{
\!\! 1\!\\
\!\!\!\!-1\!\\
\!\!\!\!-1\!\\
\!\! 0\!\\
\!\! 0\!\\
\!\! 0\!\\
\!\! 0\!\\
\!\! 0\!\\
\!\! 0\!\\
\!\! 0\!\\
\!\! 0\!\\
\!\! 0\!}
+ \rmat{
\!\! 0\!\\
\!\! 0\!\\
\!\! 0\!\\
\!\! 0\!\\
\!\! 0\!\\
\!\! 0\!\\
\!\! 0\!\\
\!\! 0\!\\
\!\! 0\!\\
\!\! 0\!\\
\!\!\!\! -\tau_{b}\!\\
\!\! \tau_{b}\!} .
\label{eq:generalSol}
\ee
In writing Eq.~\eqref{eq:generalSol}, we have chosen a specific basis in which the first basis vector (corresponding to the parameter $\alpha$) is ${\cal B} - {\cal L}$, the second (corresponding to $\beta$) is hypercharge.
The parameters $\tau_b$ are related to the exponents $u'_b$ of the $Z_N$ symmetry by
\be
u'_b=\tau_b \md N \ .
\ee
Furthermore, our basis is suitable to discuss the $[SU(2)_L]^2 - U(1)$ anomaly condition easily.
From 
\be
A_{221'}~:~ N_f (3 z[Q] + z[L]) + N_H (z[H_d] + z[H_u]) + A_{221'}^{\rm{exotic}} = 0 \label{eq:A2} 
\ee
we see that the parameters $\alpha$, $\beta$, $\gamma$, and $\tau_b$ do not enter the anomaly condition.
Plugging in the $U(1)$ charges of Eq.~\eqref{eq:generalSol}, we obtain
\be
\delta ~=~ A_{221'}^{\rm{exotic}} \ .
\ee
In the case where there are no exotic states which are charged under $SU(2)_L$, the parameter $\delta$ must therefore vanish due to the $[SU(2)_L]^2 - U(1)$ anomaly condition.
Of course, to be free from gauge anomaly, the other anomaly conditions should
also be satisfied with a specified particle spectrum.
To be as general as possible we do not consider these full gauge anomaly conditions in this paper.
However, see Refs.~\cite{Cvetic:1997ky,Cheng:1998nb,Ma:2002tc,King:2005my,Lee:2007fw,Ma:2008wq} for some examples.

Note that Eq.~\eqref{eq:generalSol} is a generalization of the discussion presented in Refs.~\cite{Lee:2007fw,Lee:2007qx} where $\delta=p=\util=\dtil=\etil=0$.
This general charge assignment is consistent with the following well-known fact: assuming (i)~Yukawa couplings with $\util=\dtil=\etil=\ntil =0$,
(ii)~no SM-charged particles other than quarks and leptons,
(iii)~vanishing of the mixed anomalies $[SU(3)_C]^2 - U(1)$ (yielding $p=-1$, see discussion in Ref.~\cite{Lee:2007fw}, for example) and $[SU(2)_L]^2 - U(1)$ (yielding $\delta = 0$),
the most general generation independent $U(1)$ which can be defined on the
quarks and leptons is a superposition of $U(1)_{B-L}$ and $U(1)_Y$, the
first and the second basis vector of Eq.~\eqref{eq:generalSol} (see also Refs.~\cite{Font:1989ai,Chamseddine:1995rs}).
Relaxing these conditions would allow different $U(1)$ symmetries.
 
Disregarding $\tau_b$, the parameters $\alpha$, $\beta$, $\gamma$, and $\delta$ can be written in terms of the $U(1)$ charges as
\be
\alpha = z[H_d]-z[L]\ , ~\quad \beta = -2z[H_d]\ , ~\quad \gamma=z[S]\ , ~\quad
\delta = -N_f (3z[Q]+z[L]) + N_H (1+p) z[S] \ . \label{eq:parameters}
\ee
In a normalization in which all $U(1)$ charges are integer, the above four
parameters (as well as $\tau_b$) are automatically also integer.
Note that the contribution of $\delta$ can be absorbed effectively in the
number of Higgs doublet pairs. However, it is not guaranteed in general that
$N_H^{\rm{eff}}$ would remain integer.

Eq.~\eqref{eq:generalSol} is useful to obtain general $U(1)$ charges in various limits.
For example, assuming $\util = \dtil = \etil = \ntil = 0$, the quark-phobic case ($z[Q] = z[U^c] = z[D^c] = 0$) requires $p=-1$, $\beta = 0$, $\delta = N_f \alpha$.\footnote{See Appendix \ref{sec:leptonic} for further discussion related to DAMA/PAMELA results.}
The lepton charges in this case are then given by
\be
z[L] = -\alpha \ , \quad z[E^c] = \alpha \ , \quad z[N^c] = \alpha \ .
\ee
The lepto-phobic case ($z[L] = z[E^c] = 0$) requires $\alpha = 0$, $\beta = 0$.
The quark charges in this case are then given by
\be
z[Q] = \frac{N_H (1+p) \gamma - \delta}{3 N_f} \, , ~~~
z[U^c] = \frac{(3 N_f - N_H) (1+p) \gamma + \delta}{3 N_f} \, , ~~~ 
z[D^c] = -\frac{N_H (1+p) \gamma - \delta}{3 N_f} \, .
\ee
Depending on value of $p$, we can categorize the models.
Especially the $p=0$ case can solve the $\mu$-problem by generating the effective $\mu$ parameter as
\be
\mu = {\hat \mu} \svev\ .
\ee
This is one of the most interesting cases for phenomenology, since the new
gauge boson $Z'$ and the exotic colored particles which are necessary to
cancel the $[SU(3)]^2_C-U(1)$ anomaly, are at the $\mu$ (TeV) scale, which can be explored by the LHC.
A TeV scale $Z'$ has implications also in cosmology such as providing a venue so that the right-handed sneutrino LSP dark matter candidate or the LUP dark matter candidate can be a thermal dark matter candidate through the $Z'$ resonance \cite{Lee:2007mt,Hur:2007ur}.
See Ref.~\cite{Langacker:2008yv} for a review of this model.
It might appear that this type of $U(1)$ cannot have matter parity ($R$-parity) as its residual discrete symmetry, but there are ways to achieve this (see Appendix~\ref{sec:RN}).

\section{Construction of the ${\bs{U(1)}}$ charges for a given discrete symmetry}
\label{sec:construction}
We discuss how to construct the most general $U(1)$ charges, which have a given discrete symmetry as its residual symmetry.
The SM-charged exotics are highly model-dependent and they may be obtained by scanning (see e.g. Refs.~\cite{Lee:2007fw,Lee:2007qx}).
Here, we limit ourselves only to the MSSM particles and the SM-singlet exotics ($X$, $T$).
The specific discrete symmetries we want to cover in this paper are listed in Table~\ref{tab:examples}. 
An overall sign change does not affect the discrete symmetry.

The general $U(1)$ charges, before any discrete symmetry is assumed, are given in Eq.~\eqref{eq:generalSol}.
Integer normalization is achieved through the coefficient $\alpha$, $\beta$, $\gamma$, and $\delta$.
Then, $N$ of $Z_N$ is determined by $z[S]$ fixing also the parameter $\gamma = N$ as shown in Eq.~\eqref{eq:parameters}.
Since invariance under a hypercharge transformation is implicitly assumed throughout the paper, the hypercharge column (with coefficient $\beta$) of Eq.~\eqref{eq:generalSol} has no effect on the discrete symmetry.
However, in order to obtain integer $U(1)$ charges, $\beta$ must be chosen in a particular way.

As a general procedure, we suggest the  following:
\begin{itemize}
\item[(i)] Take $\gamma = N$ of $Z_N$. 
\item[(ii)] Identify some terms which are allowed by the given discrete symmetry as well as the SM gauge group. 
\item[(iii)] Extract an additional condition about the $U(1)$ charges from these allowed terms (MSSM sector only). 
\item[(iv)] Using this additional relation, obtain the $U(1)$ charges from Eq.~\eqref{eq:generalSol}, the most general $U(1)$ charge assignments before imposing any particular discrete symmetry. 
\item[(v)] Require the $U(1)$ charges to be integer. 
\end{itemize}
The resulting set of equation is the most general $U(1)$ solution that contains the given discrete symmetry, up to arbitrary hypercharge shift and scaling.
We illustrate our method on three examples: $B_3$, $B_3 \times U_2$, and $B_3\times R_2$.

\TABLE{
\caption{The discrete charges of $B_3$, $B_3 \times \prod_{a,b} U^{u_a}_{a,2} U^{\prime\,{u'_b}}_{b,2}$, and $B_3 \times R_2$.
  Since $R_N=B_NL_N$, the latter symmetry can be expressed as $B_3\times R_2 =B_6^{5} L_6^3= B_6^{-1} L_6^3$ from which one can easily calculate the discrete charges in terms of $\mathcal B$, $\mathcal L$, and $y$.  
\label{tab:examples}}
{\small
\begin{tabular}{|l||r|r|r|r|r|r|r|r|r|r|r|l|
}
\hline
~symmetry~   & $\!Q\!$   & $\!\!U^c\!\!$ & $\!\!D^c\!\!$ & $\!L\!$   & $\!\!E^c\!\!$ & $\!\!N^c\!\!$ & $\!\!H_u\!\!$ & $\!\!H_d\!\!$ &
$\!X_c\!$ & $\!T^{}_d\!$ & $\!T_d^c\!\!$   & ~meaning of $q$   
\\
\hline
\hline
$\!\!B_3$          & $\!0\!$   & $\!\!\!-1\!$  & $\!1\!$   & $\!\!\!-1\!$  & $\!2\!$   & $\!0\!$   & $\!1\!$   & $\!\!\!-1\!$
& $\!0\!$ & $\!0\!$ & $\!0\!$     & $ \!y/3-{\cal B} $ 
\\
\hline
$\!\!B_3 \!\times\! \prod_{a,b} U^{u_a}_{a,2} U^{\prime\,{u'_b}}_{b,2}\!$ &
$\!0\!$   & $\!\!\!-2\!$  & $\!2\!$   & $\!\!\!-2\!$  & $\!4\!$   & $\!0\!$
& $\!2\!$   & $\!\!\!-2\!$  & $\!\!\!-3u_c \!\!$ & $\!\!\!-3u'_d\!\!$ & $\!3u'_d\!\!$     & $\!2 y/3\!-\!2 {\cal B} \!-\! 3
u_a{\cal U}_a \!-\! 3 u'_b{\cal U}'_b\!$  \\
\hline
$\!\!B_3 \times R_2$ & $\!0\!$ & $\!1\!$ & $\!\!\!-1\!$ & $\!\!\!-2\!$ & $\!1\!$
& $\!3\!$ & $\!\!\!-1\!$ & $\!1\!$ & $\!0\!$ & $\!0\!$ & $\!0\!$ & $\!- y/3+{\cal B} -3 {\cal L}  $ 
\\
\hline
\end{tabular}
}
}

\subsection{$\bs{U(1) \to B_3}$}
Here, we will consider only the MSSM sector disregarding the hidden sector fields ($X$, $T$, $T^c$).

\begin{itemize}
\item[(i)]
$B_3$ dictates $\gamma = 3$.

\item[(ii)]
To figure out the most general $U(1)$ charge assignment that contains  $B_3$, use the fact that $B_3$ allows additional terms such as $LLE^c$, $LQD^c$, and $LH_d$.
These terms can be written in a general form in the spirit of Section~\ref{sec:formalism}.
For example, $LLE^c$ can be written as
\be
\left( \frac{S}{M} \right)^n LLE^c \ ,
\ee
where $n$ is an integer.

\item[(iii)]
This gives another condition on the $U(1)$ charge assignment, 
\be
n z[S] + 2 z[L] + z[E^c] = 0 \ ,
\ee
fixing the parameter $\alpha$ in Eq.~(\ref{eq:generalSol}),
\be
\alpha = \gamma (n - \etil) = 3 (n - \etil) \ .
\ee

\item[(iv)]
Then the general solution for the MSSM part of the $B_3$ case can be written as
\be
\lmat{
z[Q]\\
z[U^c]\\
z[D^c]\\
z[L]\\
z[E^c]\\
z[N^c]\\
z[H_u]\\
z[H_d]\\
z[S]}
=
\frac{\beta}{6} \rmat{
 1\\
-4\\
 2\\
-3\\
 6\\
 0\\
 3\\
-3\\
 0}
+ \rmat{
-\etil + n + \frac{N_H(1+p)}{N_f}\\
\etil - 3\util - n + 3(1+p) - \frac{N_H(1+p)}{N_f}\\
\etil-3\dtil - n - \frac{N_H(1+p)}{N_f}\\
3\etil - 3n\\
-6\etil + 3n\\
-3\etil-3\ntil + 3n + 3(1+p)\\
-3(1+p)\\
0\\
3}
-
\frac{\delta}{3 N_f} \rmat{
 1\\
-1\\
-1\\
 0\\
 0\\
 0\\
 0\\
 0\\
 0} \ ,
\label{eq:generalB3}
\ee
with two free parameters.

\item[(v)]
Now the $U(1)$ charges should all be integers.
Regarding the first component of Eq.~(\ref{eq:generalB3}), we therefore demand $z[Q] \equiv I_Q \in \bf Z$.
This yields 
\be
\beta = 6 (I_Q+\etil-n) - 2 \cdot \frac{ 3 N_H (1+p) - \delta }{N_f}  \ , \label{eq:beta}
\ee
and Eq.~\eqref{eq:generalB3} takes the form
\be
\lmat{
z[Q]\\
z[U^c]\\
z[D^c]\\
z[L]\\
z[E^c]\\
z[N^c]\\
z[H_u]\\
z[H_d]\\
z[S]}
=
I_Q \rmat{
 1\\
-4\\
 2\\
-3\\
 6\\
 0\\
 3\\
-3\\
 0}
+ 3 \rmat{
0\\
-\util+(n-\etil)+(1+p)\\
-\dtil-(n-\etil)\\
0\\
-n\\
-\ntil+(n-\etil)+(1+p)\\
-(n-\etil)-(1+p)\\
(n-\etil)\\
1}
+
\frac{3 N_H (1+p) - \delta}{N_f} \rmat{
 0\\
 1\\
-1\\
 1\\
-2\\
 0\\
-1\\
 1\\
 0}  .
 \label{eq:generalB3two}
\ee
Due to the requirement that all charges should be integer, the coefficient of the last column, $\frac{3 N_H (1+p) - \delta}{N_f}$, must be an integer.
As already mentioned, the hypercharge column ($I_Q$) makes no difference in fixing the discrete symmetry.
The second column (with a coefficient of $3$) cannot give any net discrete charges for a $Z_3$ symmetry.
So the last column carries all the information about the discrete symmetry.
Its coefficient must be integer, but not a multiple of 3 to yield a $Z_3$ symmetry, i.e.
\be
\frac{3 N_H (1+p) - \delta}{N_f} = 3 \cdot {\bf Z} \pm 1 \ . \label{eq:prefactor}
\ee
\end{itemize}

If Eq.~(\ref{eq:prefactor}) is satisfied, the discrete charges can be easily read off from Eq.~(\ref{eq:generalB3two}) by disregarding the hypercharge column and then applying Eq.~(\ref{eq:qF}) to the remaining two vectors.
The result is $B_3$, which becomes evident by comparing the third column to the charges of $B_3$ in Table~\ref{tab:examples}.
Note that the discrete symmetry is independent of $n$ and $\util, \dtil, \etil, \ntil$.
It is also independent of $p$ as long as Eq.~\eqref{eq:prefactor} is satisfied for a given $\delta$.

Using Eq.~\eqref{eq:parameters}, we can rewrite the condition of Eq.~\eqref{eq:prefactor} as
\be
3 z[Q] + z[L] = 3 \cdot {\bf Z} \pm 1\ ,
\ee
which forbids the operator $QQQL$ effectively. This shows that $B_3$ arises automatically as a residual discrete symmetry of the $U(1)$ if we require both:
\begin{enumerate}
\item presence of an $(S/M)^n LLE^c$ term (or any effective renormalizable ${\cal L}$ violating term),
\item absence of an $(S/M)^m QQQL$ term (for any integer $m$).
\end{enumerate}
Assuming $\delta = 0$, $N_H = 1$, and $N_f=3$, the second requirement is equivalent to the requirement of the presence of an effective $\mu$ term $(S/M)^p S H_uH_d$ with $p = 3\cdot{\bf Z}$ or $3\cdot{\bf Z}+1$.
The $p=0$ case with an effective $\mu$ term ($S H_u H_d$) can belong to this category.
In the MSSM-like case with an original $\mu$ term ($H_u H_d$), i.e. $p=-1$ case, we need nonvanishing contributions from $SU(2)_L$ exotics ($\delta \ne 0$) in order to have $B_3$ as a residual discrete symmetry.

The discrete charges are given by $q = q_B  = -{\cal B} + y/3 \md 3$.
Since the hypercharge is conserved by itself, the quantity which is conserved by $B_3$ is ${\cal B} \md 3$, dictating the selection rule
\be
\Delta {\cal B} = 0 \md 3 \ .
\ee
Hence, proton decay ($\Delta B = 1$) and neutron-antineutron oscillation ($\Delta B = 2$) are absolutely forbidden by the selection rule of $B_3$ \cite{Castano:1994ec}.

Unless $R$-parity is separately imposed, this is an $R$-parity violating model.
The violation of $R$-parity implies distinguishable phenomenology.
See Refs.~\cite{Barger:1989rk,Barbieri:1989vb,Godbole:1992fb,Dreiner:2006xw,Bernhardt:2008mz,Bernhardt:2008jz} for some implications of the $R$-parity violation, for example.
The proton is still protected by $B_3$ even better than by $R$-parity \cite{Weinberg:1981wj}.
The dark matter issue still needs to be addressed.

\subsection{${\bs{U(1) \to Z_6 = B_3 \times U_2}}$}
\label{sec:B3U2}
Here we will consider the $B_3$ symmetry for the MSSM sector, augmented with $U$-parity ($U_2 = \prod_{a,b} U^{u_a}_{a,2} U^{\prime\,{u'_b}}_{b,2}$) for the hidden sector.\footnote{See Ref.~\cite{Lee:2008pc} for a special case of only Majorana hidden sector fields.}

\begin{itemize}
\item[(i)]
$Z_6$ fixes $\gamma = 6$.

\item[(ii)]
In the MSSM sector, the $LLE^c$ term can be written as
\be
\left( \frac{S}{M} \right)^n LLE^c \ ,
\ee
where $n$ is an integer.

In the hidden sector, the $X_aX_a$ and $T_bT_b^c$ terms read
\be
\left( \frac{S}{M} \right)^{{\tilde x}_a} S X_a X_a \ ,\qquad \left(\frac{S}{M} \right)^{{\tilde t}_b} S T_b T_b^c \ ,
\label{eq:hiddenmasses}
\ee
where ${\tilde x}_a$, ${\tilde t}_b$ are integers.

\item[(iii)]
For the MSSM sector we obtain
\be
n z[S] + 2 z[L] + z[E^c] = 0 \ ,
\ee
yielding the condition
\be
\alpha = \gamma (n - \etil) = 6 (n - \etil) \ .
\ee
The hidden sector mass terms in Eq.~\eqref{eq:hiddenmasses} do not give any additional constraints on the general solution because we already used these to derive Eq.~\eqref{eq:generalSol}.

\item[(iv,v)]
Demanding $I_Q \equiv z[Q]$ to be an integer, $\beta$ is given by
\be
\beta = 6 (I_Q+2\etil-2n) - \frac{2}{N_f} \left( 6 N_H (1+p) - \delta \right) , \label{eq:beta2}
\ee
and Eq.~\eqref{eq:generalSol} takes the form
\be
\lmat{
\!\!z[Q]\!\!\\
\!\!z[U^c]\!\!\\
\!\!z[D^c]\!\!\\
\!\!z[L]\!\!\\
\!\!z[E^c]\!\!\\
\!\!z[N^c]\!\!\\
\!\!z[H_u]\!\!\\
\!\!z[H_d]\!\!\\
\!\!z[S]\!\!\\
\!\!z[X_a]\!\!\\
\!\!z[T_b]\!\!\\
\!\!z[T_b^c]\!\!}
\!\!=
I_Q \rmat{
\!\! 1\!\\
\!\!\!\!-4\!\\
\!\! 2\!\\
\!\!\!\!-3\!\\
\!\! 6\!\\
\!\! 0\!\\
\!\! 3\!\\
\!\!\!\!-3\!\\
\!\! 0\!\\
\!\! 0\!\\
\!\! 0\!\\
\!\! 0\!}
+ 6 \rmat{
0\!\\
\!\!\!\!-\util+(n\!-\!\etil)\!+\!(1\!+\!p)\!\\
\!\!\!\!-\dtil-(n\!-\!\etil)\!\\
0\!\\
-n\!\\
\!\!\!\!-\ntil+(n\!-\!\etil)\!+\!(1\!+\!p)\!\\
-(n\!-\!\etil)\!-\!(1\!+\!p)\!\\
(n\!-\!\etil)\!\\
1\!\\
0\!\\
0\!\\
-(1+{\tilde t}_b)\!}
+
\frac{6 N_H (1\!+\!p) \!-\! \delta}{N_f} \rmat{
\!\! 0\!\\
\!\! 1\!\\
\!\!\!\!-1\!\\
\!\! 1\!\\
\!\!\!\!-2\!\\
\!\! 0\!\\
\!\!\!\!-1\!\\
\!\! 1\!\\
\!\! 0\!\\
\!\! 0\!\\
\!\! 0\!\\
\!\! 0\!}
 +
\rmat{
 \!\!0\!\\
 \!\!0\!\\
 \!\!0\!\\
 \!\!0\!\\
 \!\!0\!\\
 \!\!0\!\\
 \!\!0\!\\
 \!\!0\!\\
 \!\!0\!\\
\!\!\!\!-3(1\!+\!{\tilde x}_a)\!\\
\!\!\!\!-\tau_b\! \\
\!\! \tau_b\!}  \!\! .
 \label{eq:generalB3U2two}
\ee
Again the first two columns have no effect on the discrete symmetry, which is therefore only determined by the coefficient of the third vector, $\frac{6 N_H (1+p) - \delta}{N_f}$, as well as the parameters ${\tilde x}_a$ and $\tau_b$.
The former defines the discrete charges of the MSSM fields while the latter two fix those of the hidden sector particles.
As we are looking for the case with $B_3$ among the MSSM fields, we must require (see
Table~\ref{tab:examples}) 
\be
\frac{6 N_H (1+p) - \delta}{N_f} = 6 \cdot {\bf Z} \pm 2 \ .\label{BUcondi}
\ee
On the other hand, the $Z_2$ symmetry ($U$-parity) of the hidden sector necessitates 
\be
\tau_b ~=~3 \cdot {\bf Z} \ ,
\ee
whereas ${\tilde x}_a$ remains unconstrained.
It is worth noting that the discrete symmetry is independent of ${\tilde t}_b$.
\end{itemize}

With only one Majorana $X$ and one Dirac $T$ particle in the hidden sector, one can have three different nontrivial  scenarios:
\begin{itemize}
\item $X$ is odd and $T$ is even under $U$-parity, i.e. $u=1$ and $u'=0$.
This requires ${\tilde x} = 2 \cdot  {\bf Z}$ and $\tau = 6 \cdot {\bf Z}$.
Reversely, if $\left(\frac{S}{M}\right)^{\tilde x} SXX$ with $\tilde x = 0, 2, 4, \cdots$ exists, the hidden field $X$ automatically has odd $U$-parity.
\item $X$ is even and $T$ is odd under $U$-parity, i.e. $u=0$ and $u'=1$.
Such a situation requires ${\tilde x} = 2 \cdot {\bf Z} +1$ and $\tau = 6 \cdot {\bf Z} + 3$.
The exponent $\tilde t$ in the mass term $\left(\frac{S}{M}\right)^{\tilde t}TT^c$ does not enter the discussion of the discrete symmetry.
\item $X$ is odd and $T$ is odd under $U$-parity, i.e. $u=1$ and $u'=1$.
Here we need ${\tilde x} = 2 \cdot {\bf Z}$ and $\tau = 6 \cdot {\bf Z} + 3$.
In this case, the lighter of the two particles will be stable due to $U$-parity.
\end{itemize}

Using Eq.~\eqref{eq:parameters}, we can rewrite the condition of Eq.~(\ref{BUcondi})  as
\be
3 z[Q] + z[L] = 6 \cdot {\bf Z} \pm 2 \ ,
\ee
which forbids the operator $QQQL$ effectively.
Therefore, a symmetry of type $B_3 \times \prod_{a} U^{u_a}_{a,2}$ arises automatically as a residual discrete symmetry of the $U(1)$ if we require: 
\begin{enumerate}
\item presence of an $(S/M)^n LLE^c$ term (or any effective renormalizable ${\cal L}$ violating term),
\item absence of an $(S/M)^m QQQL$ term (for any integer $m$),
\item presence of an $\left(\frac{S}{M}\right)^{{\tilde x}_a} SX_aX_a$ term with ${\tilde x}_a = 0, 2, 4, \cdots$, resulting in $X_a$ being odd under $U$-parity.
\end{enumerate}
Unfortunately, the case where $T_b$ has odd $U$-parity cannot be discussed in terms of requiring the presence or absence of some effective operators as discussed above. 

The discrete charges are given by $q = 2 q_B + 3 u_a q_{U_a}+ 3 u'_b q_{U'_b} \md 6 = -2 {\cal B} + 2y/3 - 3 u_a{\cal U}_a - 3 u'_b{\cal U}'_b \md 6$.
Since the hypercharge is conserved by itself, the quantity which is conserved by $B_3$ is ${\cal B} \md 3$  and the one conserved by $U$-parity is $u_a{\cal U}_a + u'_b{\cal U}'_b\md 2$, dictating the selection rules
\be
\Delta {\cal B} = 0 \md 3 \ , \qquad \Delta (u_a{\cal U}_a + u'_b{\cal U}'_b)
= 0 \md 2 \ ,
\ee
which prevents the proton and the LUP from decaying.
Therefore, $R$-parity is not necessary to address the stability of the proton and the dark matter candidate.

\subsection{${\bs{U(1) \to Z_6 = B_3 \times R_2}}$}
\label{sec:B3R2}
Here we will consider the $B_3 \times R_2$ symmetry for the MSSM sector without any hidden sector fields.\footnote{See Ref.~\cite{Dreiner:2005rd,Dreiner:2007vp} for details about this symmetry.}
As we can check with Table~\ref{tab:examples}, this symmetry allows $\left( \frac{S}{M} \right)^n (H_u L)^2$ which can provide an additional condition.
Applying the general procedure, we find
\be
\lmat{
z[Q]\\
z[U^c]\\
z[D^c]\\
z[L]\\
z[E^c]\\
z[N^c]\\
z[H_u]\\
z[H_d]\\
z[S]}
=
I_Q \rmat{
 1\\
-4\\
 2\\
-3\\
 6\\
 0\\
 3\\
-3\\
 0}
+ 6 \rmat{
0\\
-\util\\
-\dtil+(1+p)\\
0\\
-\etil+(1+p)\\
-\ntil\\
0\\
-(1+p)\\
1}
+
\frac{6 N_H (1+p) - \delta}{N_f} \rmat{
 0\\
 1\\
-1\\
 1\\
-2\\
 0\\
-1\\
 1\\
 0}
 +
 3 n \rmat{
 0\\
 1\\
-1\\
 0\\
-1\\
 1\\
-1\\
 1\\
 0} \ .
 \label{eq:generalB3R2}
\ee
To have the $B_3 \times R_2$, the last two vectors should results in the discrete charges of Table~\ref{tab:examples}.
Then we need
\be
\frac{6 N_H (1+p) - \delta}{N_f} = 6 \cdot {\bf Z} \mp 2 \ , \qquad 3 n = 6
\cdot {\bf Z} + 3 \ ,
\label{eq:B3R2condition}
\ee
where the second equation requires $n$ to be an odd integer.

\section{Summary and conclusions}
In this paper, we systematically studied the residual discrete symmetry of an extra Abelian gauge symmetry, which may interact with both the MSSM sector and the hidden sector.
Despite a common gauge origin, the discrete symmetry can have important implications separately for the observable and the hidden sector, such as the stability of the proton {\it and} dark matter.
We provided the most general framework to discuss such a symmetry including Majorana type and Dirac type hidden sector fields.

We also argued how to construct the most general $U(1)$ symmetry for the MSSM sector and hidden sector for a given discrete symmetry, illustrating our procedure for several examples.
Our results should be useful for $U(1)$ model building.
For example, in order to make sure the proton and the Majorana hidden sector dark matter candidate are stable in the absence of $R$-parity, one can, in a minimal framework with $\delta=0$, $N_H=1$, and $N_f=3$, just require (i) $S H_u H_d$ (i.e. the effective $\mu$ term that solves the $\mu$-problem with the $U(1)$ gauge symmetry), (ii) $LLE^c$  (a renormalizable $\cal L$ violating term), and (iii) $SXX$  (a mass term for the Majorana hidden sector field $X$).
Then, $B_3$ and $U$-parity are automatically invoked in the MSSM and the hidden sector, respectively, as a residual discrete symmetry of the common $U(1)$ gauge symmetry (in the form of $Z_6 = B_3 \times U_2$). Their selection rules ensure absolute stability of the proton and the LUP dark matter. 

In Appendix \ref{sec:RN}, we investigated the cases in which the $U(1)$ gauge symmetry that solves the $\mu$-problem can contain matter parity (equivalent to $R$-parity) as a residual discrete symmetry.
This can provide a useful framework for $R$-parity conserving $U(1)$ extended supersymmetric models, without imposing a separate $R$-parity.

In Appendix \ref{sec:leptonic}, we made a comment on the relation between the discrete symmetry and the leptonically interacting LUP dark matter candidate, which has been recently focused on due to the possibility of the simultaneous explanation of the DAMA modulation and the PAMELA results.

\appendix
\section{${\bs U(1)}$ gauge origin of ${\bs R}$-parity and the ${\bs \mu}$-problem solution}
\label{sec:RN}
In this appendix, we investigate the conditions under which the $R_2$ matter parity (equivalent to the $R$-parity) can emerge as a residual discrete symmetry of the extra $U(1)$ gauge symmetry. 
The general $U(1)$ charges of the MSSM sector and the hidden sector are given in Eq.~(\ref{eq:generalSol}).
In order to unveil the discrete symmetry, let us introduce a new parameter $\beta'$ which is related to the original parameters $\alpha$, $\beta$, $\gamma$, $\delta$ by
\be
\frac{\beta}{6} = \frac{\beta'}{6} -\frac{\alpha}{3} - \frac{\gamma}{3N_f}N_H(1+p) +\frac{\delta}{3N_f} \ .
\ee
Using this definition, Eq.~(\ref{eq:generalSol}) can be rewritten to separate the columns into those which do and which do not affect the discrete symmetry (in the $R_N$ and $B_N$ basis) among the MSSM fields.
In the following, we do not consider the the hidden sector fields, which are irrelevant to our discussion.
\be
\lmat{
z[Q]\\
z[U^c]\\
z[D^c]\\
z[L]\\
z[E^c]\\
z[N^c]\\
z[H_u]\\
z[H_d]\\
z[S]}
=
-\alpha \underbrace{\rmat{
 0\\
-1\\
 1\\
0\\
 1\\
-1\\
 1\\
-1\\
 0}}_{q_R}
- \frac{\gamma N_H (1+p) - \delta}{N_f}   \underbrace{\rmat{
 0\\
-1\\
 1\\
-1\\
 2\\
 0\\
 1\\
-1\\
 0}}_{q_B} +
\frac{\beta'}{6} \rmat{
 1\\
-4\\
 2\\
-3\\
 6\\
 0\\
 3\\
-3\\
 0}
+  \gamma    \rmat{
 0\\
-\util +  (1+p)  \\
-\dtil  \\
0  \\
-\etil\\
-\ntil+ (1+p)\\
- (1+p) \\
 0 \\
 1}  .
\ee
As mentioned before, the hypercharge (third) column does not influence the discrete symmetry at all since we require hypercharge shift invariance.
The fourth column has no effect on the discrete symmetry among the MSSM fields because their contributions to the $U(1)$ charges are integer multiples of $\gamma=z[S]=N$.
Therefore only the first and the second column define the discrete symmetry among the MSSM fields.
Comparing the entries of both vectors with the discrete charges of Table~\ref{tab:discrete}, we see that the first column corresponds to $q_R$ and the second to $q_B$.
Hence, the type of $Z_N^{\rm{obs}}$ symmetry depends only on the coefficients of these two vectors, namely on $\alpha$  and $\frac{\gamma N_H (1+p) - \delta}{N_f}$.
Our assumption of integer $U(1)$ charges requires both to be integer.
Note that $\frac{\beta'}{6}$ is necessarily also integer and can be replaced by $I_Q$ (see step (v) of the examples in Section~\ref{sec:construction}). 

In order to have a pure $R_N$ symmetry, the coefficient of $q_R$ should be $\pm 1 \md N$ and the coefficient of $q_B$ must vanish $\md N$, i.e.
\be
\alpha = \gamma \cdot {\bf Z} \pm 1 \ , \qquad
\frac{\gamma N_H (1+p) -\delta}{N_f}  = \gamma \cdot {\bf Z} \ .
\label{eq:RN?}
\ee

With $p=-1$, $\delta = 0$, and $\util = \dtil = \etil = \ntil = 0$, we have only a mixtures of the $U(1)_{B-L}$ and $U(1)_Y$.
It always has the pure $R_N$ symmetry as a residual discrete symmetry as long as $\alpha = \gamma \cdot {\bf Z} \pm 1$.
With $\gamma = 2$, we obtain $R_2$ parity.
Relaxing $\util = \dtil = \etil = \ntil = 0$ does not change the discrete symmetry.

In order to solve the $\mu$-problem with a TeV scale $U(1)$ gauge symmetry, however, we should take $p=0$ (see Section~\ref{sec:formalism}).
With $N_f = 3$, Eq.~\eqref{eq:RN?} can be written as
\be
\delta = \gamma (N_H - 3 \cdot {\bf Z}) \ .
\ee
Therefore we need additional $SU(2)_L$ exotic fields in the form of $N_H =3 \cdot {\bf Z}$ generations of Higgs pairs or some exotic doublet contribution $\delta$, in order to have $Z_N^{\rm obs} = R_N$ while solving the $\mu$-problem with a common $U(1)$ gauge symmetry.

However, it should be mentioned that there is another way for the $U(1)$ to be a solution to the $\mu$-problem while having matter parity as a residual discrete symmetry, which may not require additional $SU(2)_L$ charged particles.
If the total discrete symmetry in the MSSM sector has $R_2$ as a {\it part} of it, i.e. $Z_{N}^{\rm obs} = R_2 \times Z_{N/2}$ (where $2$ and $N/2$ are coprime), both $R_2$ and $Z_{N/2}$ will be conserved independently.
For instance, consider $B_3 \times R_2$ as the $U(1)$ residual discrete symmetry as in Section~\ref{sec:B3R2}.
With $N_f=3$ and $N_H = 1$, and no $SU(2)_L$ exotics ($\delta = 0$),
Eq.~\eqref{eq:B3R2condition} gives $1+p = 3 \cdot {\bf Z} \mp 1$, which allows $p=0$ to solve the $\mu$-problem.
Hence, one $U(1)$ gauge symmetry can be the common source of the $\mu$-problem solution as well as $R$-parity.

\section{Leptonically coupling dark matter}
\label{sec:leptonic}
It is worth to note that various coupling limits are still compatible with discrete symmetries.
For instance, consider the $B_3 \times U_2$ we studied in Section~\ref{sec:B3U2}.
The quark-phobic case ($z[Q] = z[U^c] = z[D^c] = 0$) requires $\delta = 6 (N_f (\dtil - \etil + n) + N_H (1+p) )$ and $\util + \dtil = (1+p)$, and $I_Q = 0$.
The lepton charges in this case, up to arbitrary scaling, are
\be
z[L] = -6 (\dtil - \etil + n) \ , \quad
z[E^c] = 6 (2\dtil - 2\etil + n) \ , \quad
z[N^c] = 6 (- \etil - \ntil + n + (1+p) ) \ . \label{eq:quarkphobic}
\ee

In particular, a dark matter candidate that interacts with only leptons has been paid good attention since it may be able to explain the DAMA annual modulation without making conflict with other direct detection experiments \cite{Bernabei:2007gr}.
This kind of dark matter would be consistent with the property that can naturally explain the PAMELA results (for example, see Ref.~\cite{Cirelli:2008pk}).
PAMELA showed a significant positron excess \cite{Adriani:2008zr} but no deviation in the proton/antiproton data \cite{Adriani:2008zq}.
Since the LUP dark matter, which is stable under the $U$-parity, interacts with the gauge boson $Z'$ of the $U(1)$ gauge symmetry, the quark-phobic case can satisfy this property.
See Ref.~\cite{Fox:2008kb} for an illustration how such a dark matter can explain the DAMA and PAMELA results.\footnote{There are other scenarios to explain the PAMELA data with an extra $U(1)$ gauge symmetry. For example, see Ref.~\cite{Chen:2008yi} for an illustration how a hidden $U(1)$ gauge symmetry, which interacts with the SM particles via kinetic mixing can explain such a signature.}

As $B_3 \times U_2$ is compatible with the quark-phobic case, the existence of such a dark matter may not only explain the DAMA and PAMELA results but also suggests why the proton and dark matter are stable without introducing separate parities.
This scenario may be tested, for example, by the precise measurement of $Z'$ coupling to leptons and comparison with Eq.~\eqref{eq:quarkphobic}.

\begin{acknowledgments}
We are grateful to K. Matchev, S. Nasri and J. Yoo for useful discussions.
HL is supported by the Department of Energy under grants
DE-FG02-97ER41029 and DE-FG03-94ER40837.
CL is supported by University of Florida through the Institute for Fundamental Theory as well as the STFC rolling grant ST/G000557/1.
\end{acknowledgments}


\end{document}